\documentstyle[12pt,epsf]{article}
 \newcommand{\insertplot}[5]{\begin{figure}
 \hfill\hbox to 0.05in{\vbox to #5in{\vfill
 \inputplot{#1}{#4}{#5}}\hfill}
 \hfill\vspace{-.1in}
 \caption{#2}\label{#3}
 \end{figure}}
 \newcommand{\inputplot}[3]{
 \special{ps: plotfile #1}
}
\newcommand{\vphi}{\varphi}

\begin{document}

\title{
Non-Abelian Black Holes\\ with Magnetic Dipole Hair}
\vspace{1.5truecm}
\author{
{\bf Burkhard Kleihaus}\\
Department of Mathematical Physics, University College, Dublin,\\
Belfield, Dublin 4, Ireland\\
{\bf Jutta Kunz}\\
Fachbereich Physik, Universit\"at Oldenburg, Postfach 2503\\
D-26111 Oldenburg, Germany}

\vspace{1.5truecm}

\date{\today}

\maketitle
\vspace{1.0truecm}

\begin{abstract}
We construct static axially symmetric black holes
in SU(2) Einstein-Yang-Mills-Higgs theory.
Located inbetween a monopole-antimonopole pair, 
these black holes possess magnetic dipole hair.
The difference of their mass and their horizon mass
equals the mass of the regular monopole-antimonopole solution,
as expected from the isolated horizon framework.
\end{abstract}
\vfill
\noindent {Preprint hep-th/0008034} \hfill\break
\vfill\eject

\section{Introduction}
In Einstein-Maxwell (EM) theory, black holes are completely determined
by their mass, their charge and their angular momentum,
i.e.~black holes have no hair \cite{nohair}.
Furthermore, Israel's theorem of EM theory states, that
static black holes are spherically symmetric.
The ``no-hair'' theorem holds no longer in
theories with non-abelian fields. Such theories
possess whole sequences of neutral static spherically symmetric 
black hole solutions \cite{review}.
Furthermore, there are static black hole solutions
with only axial symmetry \cite{kk},
as well as black hole solutions with only discrete symmetries \cite{ewein},
so Israel's theorem holds neither in the presence of non-abelian fields.

The hairy black hole solutions are asymptotically flat
and possess a regular event horizon \cite{review}.
Taking the radius of the event horizon to zero,
globally regular particle-like solutions emerge.
Recently, in the isolated horizon framework
an intriguing relation between the
mass of hairy black hole solutions
and the mass of the corresponding regular solutions
was found in Einstein-Yang-Mills (EYM) theory
\cite{ashtekar,sudar}.
Also, a modified uniqueness conjecture was presented
\cite{sudar}.

SU(2) Einstein-Yang-Mills-Higgs (EYMH) theory,
in particular, allows for a rich variety of
globally regular particle-like solutions, such as
gravitating monopoles \cite{gmono} and multimonopoles \cite{hkk}
as well as gravitating monopole-antimonopole pairs \cite{map2}.
Associated with the gravitating monopole solutions are
non-abelian black hole solutions with ``magnetic monopole hair'',
existing for not too large values of the event horizon
\cite{gmono}. Likewise, 
the gravitating multimonopole solutions give rise to
magnetically charged non-abelian black hole solutions with hair \cite{ewein}.

Here we show, that the 
regular gravitating monopole-antimonopole pair (MAP) solutions \cite{map2} 
are similarly associated with hairy black hole solutions.
Immersing a black hole symmetrically between the monopole and antimonopole
results in a static axially symmetric black hole solution carrying
``magnetic dipole hair''. 
We further show that the difference of the mass 
and the horizon mass of these hairy black holes 
equals the mass of the regular MAP solution,
as expected from the isolated horizon formalism \cite{ashtekar}.

\section{Ansatz}
We consider SU(2) EYMH theory with action
\begin{equation}
S=\int \left ( \frac{R}{16\pi G} 
+ \frac{1}{2e} {\rm Tr} (F_{\mu\nu} F^{\mu\nu})
-\frac{1}{4}{\rm Tr}(D_\mu \Phi D^\mu \Phi)
  \right ) \sqrt{-g} d^4x
\ ,   \end{equation}
and with Newton's constant $G$,
Yang-Mills coupling constant $e$, 
Higgs vacuum expectation value $\eta$,
and vanishing Higgs self-coupling.
Variation with respect to the metric and the matter fields
leads to the Einstein equations and the field equations,
respectively.

The static axially symmetric black hole solutions with ``magnetic
dipole hair'' are obtained in isotropic coordinates with metric \cite{kk}
\begin{equation}
ds^2=
  - f dt^2 +  \frac{m}{f} \left( d r^2+ r^2d\theta^2 \right)
           +  \frac{l}{f} r^2\sin^2\theta d\vphi^2
\ , \label{metric} \end{equation}
where $f$, $m$ and $l$ are only functions of $r$ and $\theta$.
The MAP ansatz reads for the purely magnetic
gauge field ($A_0=0$) \cite{map,map2}
\begin{equation}
A_\mu dx^\mu = 
\frac{1}{2e}\left\{
\left(\frac{H_1}{r}dr+2(1-H_2)d\theta\right)\tau_\vphi
-2\sin\theta\left(H_3 \tau_r^{(2)}+(1-H_4) 
\tau_\theta^{(2)}\right)d\vphi
\right\}
\label{ansatz} \end{equation}
and for the Higgs field
\begin{equation}
\Phi= \left(\Phi_1 \tau_r^{(2)}+\Phi_2 \tau_\theta^{(2)}\right)
\ , \end{equation}
with $su(2)$ matrices (composed of the standard Pauli matrices $\tau_i$)
\begin{eqnarray}
& \tau_r^{(2)} = \sin 2\theta \tau_\rho +\cos 2\theta\tau_3 \ ,
\ \ \ 
\tau_\theta^{(2)}  = \cos 2\theta \tau_\rho -\sin 2\theta\tau_3 \ , &
\nonumber\\
& \tau_\rho  =  \cos \vphi \tau_1+ \sin \vphi\tau_2\ ,
\ \ \ 
\tau_\vphi   =  -\sin \vphi \tau_1+ \cos \vphi\tau_2\ .
\end{eqnarray}
The four gauge field functions $H_i$ and the two Higgs field functions
$\Phi_i$ depend only on $r$ and $\theta$.
Consistent with this ansatz,
the general set of equations of motion is reduced to a set of
nine elliptic partial differential equations.
With respect to the residual gauge degree of freedom \cite{rr,kk,map} 
we choose the gauge condition 
$r\partial_r H_1-2\partial_\theta H_2 =0$ \cite{map,map2}.

\section{Boundary conditions}
We consider static axially symmetric black hole solutions
with ``magnetic dipole hair'',
which are asymptotically flat, and possess a finite mass.
Their boundary conditions at infinity and along the $\rho$- and $z$-axis
are the same as those of the regular MAP solutions \cite{map2},
i.e.~at infinity the conditions are
\begin{equation}
H_1=H_2=0 \ ,
H_3=\sin \theta \ , 1-H_4=\cos\theta \ , 
 \Phi_1 = \eta \ , \Phi_2=0 \ , f=m=l=1 \ ,
\end{equation}
on the $z$-axis the functions 
$H_1, H_3, \Phi_2$ and the derivatives
$\partial_\theta H_2,\partial_\theta H_4,\partial_\theta \Phi_1 ,
\partial_\theta f, \partial_\theta m,\partial_\theta l$ have to vanish,
while on the $\rho$-axis the functions
$H_1, 1-H_4, \Phi_2$ and the derivatives
$\partial_\theta H_2,\partial_\theta H_3,\partial_\theta \Phi_1 ,
\partial_\theta f, \partial_\theta m,\partial_\theta l$ have to vanish.

In order to obtain black hole solutions with a
regular event horizon at radius $r_{\rm H}$,
we impose the boundary conditions \cite{kk}
\begin{equation}
 f=m=l=0 \ , \ \ \  
 r \partial_r \Phi_1 + H_1 \Phi_2 =0 \ , \ \ \
 r \partial_r \Phi_2 - H_1 \Phi_1 =0 \ ,  
\nonumber \end{equation}
\begin{equation}
\partial_\theta H_1 + r \partial_r H_2 = 0 \ , \ \ \
 r \partial_r H_3-H_1 H_4=0,   \ \ \ 
 r \partial_r H_4+H_1( H_3 + {\rm ctg} \theta) =0
\ , \end{equation}
where the boundary conditions on the matter field functions
result from the equations of motion.
Since one of the gauge field equations 
gives a boundary condition which coincides with the gauge fixing condition,
$ r \partial_r H_1 - \partial_\theta H_2 = 0 $,
we need to impose a further condition at the horizon
to completely fix the gauge \cite{kk}.
We choose $\partial_\theta H_1 = 0$.

From the equations of motion it follows \cite{kk}, that
the Kretschmann scalar is finite at the horizon,
and that the surface gravity $\kappa$ \cite{wald,ewein},
\begin{equation}
\kappa^2=-(1/4)g^{tt}g^{ij}(\partial_i g_{tt})(\partial_j g_{tt})
\ , \label{sg} \end{equation}
is constant, as required by the zeroth law of black hole physics.

\section{Results}
Introducing the dimensionless coordinate 
$x=r\eta e$ and the Higgs field $\phi = \Phi/\eta$,
the equations depend only on the coupling constant $\alpha$,
$\alpha^2 = 4\pi G\eta^2$.
The mass $M$ of the black hole
solutions can be obtained directly from
the total energy-momentum ``tensor'' $\tau^{\mu\nu}$
of matter and gravitation,
$M=\int \tau^{00} d^3r$ \cite{wein},
or equivalently from
$ M = - \int \left( 2 T_0^{\ 0} - T_\mu^{\ \mu} \right)
   \sqrt{-g} dr d\theta d\varphi $,
yielding the dimensionless mass $\mu/\alpha^2 = \frac{e}{4\pi\eta} M$.

Let us briefly recall the globally regular 
gravitating MAP solutions.
In the limit $\alpha \rightarrow 0$, 
the lower branch of gravitating MAP solutions 
emerges smoothly from the flat space solution \cite{map}
and ends at the critical value $\alpha_{\rm cr}^{\rm reg}=0.670$,
when gravity becomes too strong for regular 
MAP solutions to persist \cite{map2}.
However, at $\alpha_{\rm cr}^{\rm reg}$
a second branch of MAP solutions appears,
extending back to $\alpha=0$.
Having higher masses, these MAP solutions constitute the upper branch
of solutions.
On both branches,
the modulus of the Higgs field of the gravitating MAP solutions
possesses two zeros, $\pm z_0$, on the $z$-axis,
which correspond to the location of the monopole and antimonopole, 
respectively.

Let us now turn to
the corresponding black hole solutions with ``magnetic dipole hair'',
where a regular event horizon is located 
inbetween the monopole-antimonopole pair.
These solutions are obtained by solving 
the set of equations of motion numerically,
subject to the above boundary conditions \cite{xgmap}.

For a fixed value of $\alpha$, 
$0 < \alpha < \alpha_{\rm cr}^{\rm reg}$,
we obtain two branches of black hole solutions.
Imposing a regular event horizon at a small radius $x_{\rm H}$,
the lower branch of black hole solutions emerges from the 
regular lower branch MAP solution.
Along this lower branch, with increasing $x_{\rm H}$ the mass increases.
Indeed, when a maximal value of the horizon radius $x_{\rm H,max}(\alpha)$
is reached, the mass increases further with decreasing $x_{\rm H}$,
reaching a maximal value
at $x_{\rm H,cr}(\alpha)<x_{\rm H,max}(\alpha)$.
Decreasing $x_{\rm H}$ from
$x_{\rm H,cr}(\alpha)$, a second branch of solutions appears.
Along this upper branch with decreasing $x_{\rm H}$ the mass decreases,
reaching the regular upper branch MAP solution,
when $x_{\rm H} \rightarrow 0$.

We now introduce the area parameter $x_\Delta$ \cite{ashtekar,sudar},
defined via the dimensionless area of the black hole horizon $A$,
\begin{equation}
A = 2 \pi x_{\rm H}^2
\int_0^\pi  d\theta \sin \theta
\left. \frac{\sqrt{l m}}{f} \right|_{x_{\rm H}^2} 
 = 4 \pi x_\Delta^2
\ . \label{area} \end{equation}
When considered as a function of the horizon radius $x_{\rm H}$,
the area parameter $x_\Delta$
also attains its maximal value at the critical value $x_{\rm H,cr}$.
Therefore the dimensionless mass $\mu$,
when considered as a function of the area parameter $x_{\Delta}$,
attains its maximum value at the maximum value of $x_\Delta$.
Here the two branches merge, forming a spike, as seen in Fig.~1 
for $4 \pi \alpha^2 = 0.725$.

The maximal value of the horizon radius
$x_{\rm H,max}(\alpha)$ increases with decreasing
$\alpha$. 
The weaker gravity, the bigger black holes are possible.
Extrapolating to the limit $\alpha \rightarrow 0$,
we obtain the biggest possible horizon radius, $x_{\rm H}=0.1208$.

Let us now consider these black hole solutions in 
the deformed isolated horizon framework \cite{ashtekar,sudar}.
For EYM theory an intriguing relation
between the ADM mass $\mu$ of a black hole with area parameter $x_\Delta$
and the mass $\mu_{\rm reg}$ of the corresponding globally regular solution
was found \cite{ashtekar},
\begin{equation}
\mu =  \mu_{\Delta}
+\mu_{\rm reg}
\label{ash} \ , \end{equation}
where the horizon mass $\mu_{\Delta}$ is defined via 
\begin{equation}
\mu_{\Delta} = \int_0^{x_{\Delta}} \kappa(x'_{\Delta}) 
 x'_{\Delta} d x'_{\Delta}
\label{Mhor} \ , \end{equation}
with rescaled surface gravity $\kappa \rightarrow \kappa/e\eta$.

We now consider this relation for the EYMH black holes 
with ``magnetic dipole hair''.
The inverse of the surface gravity $\kappa$, 
and the integrand $\kappa(x_\Delta) x_\Delta$
of the integral for
the horizon mass, are shown in Fig.~2 for $4 \pi \alpha^2 = 0.725$
as functions of
the area parameter $x_\Delta$ for both branches of black hole solutions.

Evaluating the rhs of relation (\ref{ash}) for the black hole solutions
along both branches,
we obtain excellent agreement with the corresponding ADM masses, 
as seen in Fig.~1.
Thus our numerical results indicate, that
relation (\ref{ash}) also holds for the EYMH black hole solutions
with ``magnetic dipole hair''.
(For each branch the corresponding regular solution
is the reference point for the integration.)
In particular, the mass of the regular upper branch solution is
related to the mass of the regular lower branch solution
via the horizon mass integral, performed along both branches.

Another crucial quantity in this formalism is the
magnetic non-abelian charge of the horizon 
$P_\Delta^{\rm YM}$ \cite{ashtekar,sudar}
\begin{equation}
P_\Delta^{\rm YM} = \frac{1}{4 \pi} \oint 
 \sqrt{ \sum_i \left( F^i_{\theta\varphi} \right)^2} d\theta d\varphi
\ , \end{equation}
where the integral is over the surface of the horizon.
We show $P_\Delta^{\rm YM}$ in Fig.~3. Also shown is the
magnetic abelian charge of the horizon $P_\Delta^{M}$,
obtained analogously from the electromagnetic 't Hooft tensor 
${\cal F}_{\mu\nu}$ \cite{mono}.
Note, that the magnetic charge inside the horizon,
$P^M = \oint {\cal F}_{\theta\varphi} d\theta d\varphi$,
vanishes.
These charges are of interest also, since
in \cite{sudar} a modified uniqueness conjecture for black holes was suggested,
claiming that black holes are uniquely specified by their horizon charges.

Further details to these black hole solutions as well as
to possible excited black hole solutions will be given elsewhere.

\section{Outlook}
Having constructed black hole solutions with ``magnetic dipole hair'',
where the black hole resides at the center between the poles,
it is natural to wonder, whether one could also obtain
static axially symmetric solutions with two black holes,
each located at one of the poles, i.e.~black diholes \cite{bilbao}.
Such solutions would presumably form an unstable equilibrium
configuration of two hairy black holes.

When considering a magnetic dipole configuration 
with two black holes in Einstein-Maxwell theory,
one has either to invoke a string between the holes
or an external magnetic field to give the necessary
repulsion and avoid collapse \cite{bilbao}. 
For the non-abelian configuration, in contrast,
we expect the gauge field to provide the necessary repulsion.
Indeed, for the globally regular MAP solutions
and the above constructed black hole solutions with 
``magnetic dipole hair'', 
the gauge field provides sufficient repulsion for equilibrium.

For the single black hole with ``magnetic dipole hair''
our numerical results are in excellent agreement 
with relation (\ref{ash}), 
obtained in the isolated horizon framework.
It appears interesting to consider extension of this
relation to the conjectured black dihole configurations.

\newpage
\begin{figure}[h!]\centering
\epsfysize=11cm\mbox{\epsffile{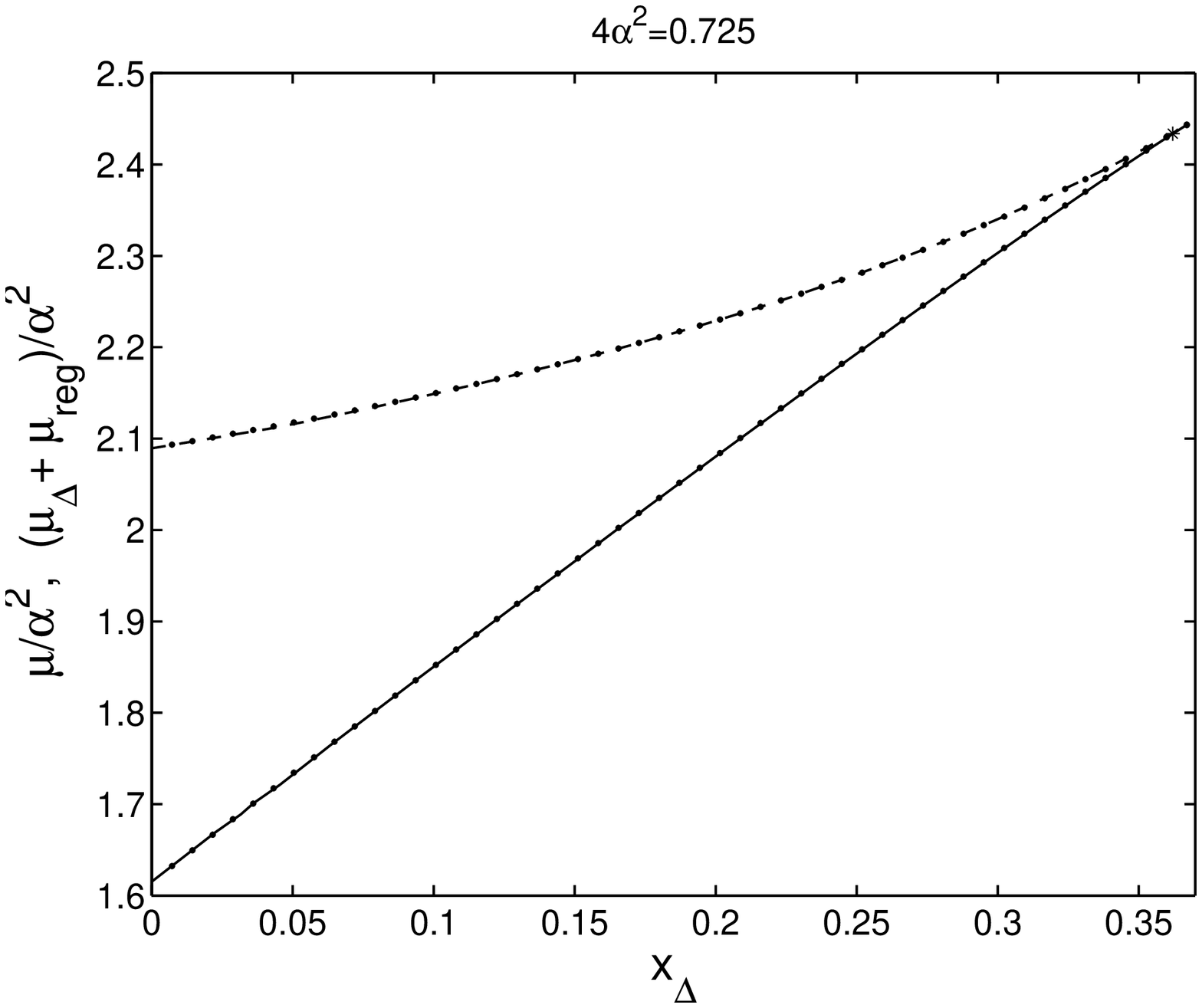}}
\caption{
The ADM mass $\mu/\alpha^2$ along the lower branch (solid)
and the upper branch (dashed)
is shown for the EYMH black hole solutions
with ``magnetic dipole hair'' as a function of the
area parameter $x_\Delta$ for coupling constant 
$4 \pi \alpha^2 = 0.725$.
The asterisk ($*$) indicates the solution with
horizon radius $x_{\rm H,max}$.
Also shown is the rhs of relation (11)
$(\mu_\Delta + \mu_{\rm reg})/\alpha^2$
(dotted).
}
\end{figure}
\newpage
\begin{figure}[h!]\centering
\epsfysize=11cm\mbox{\epsffile{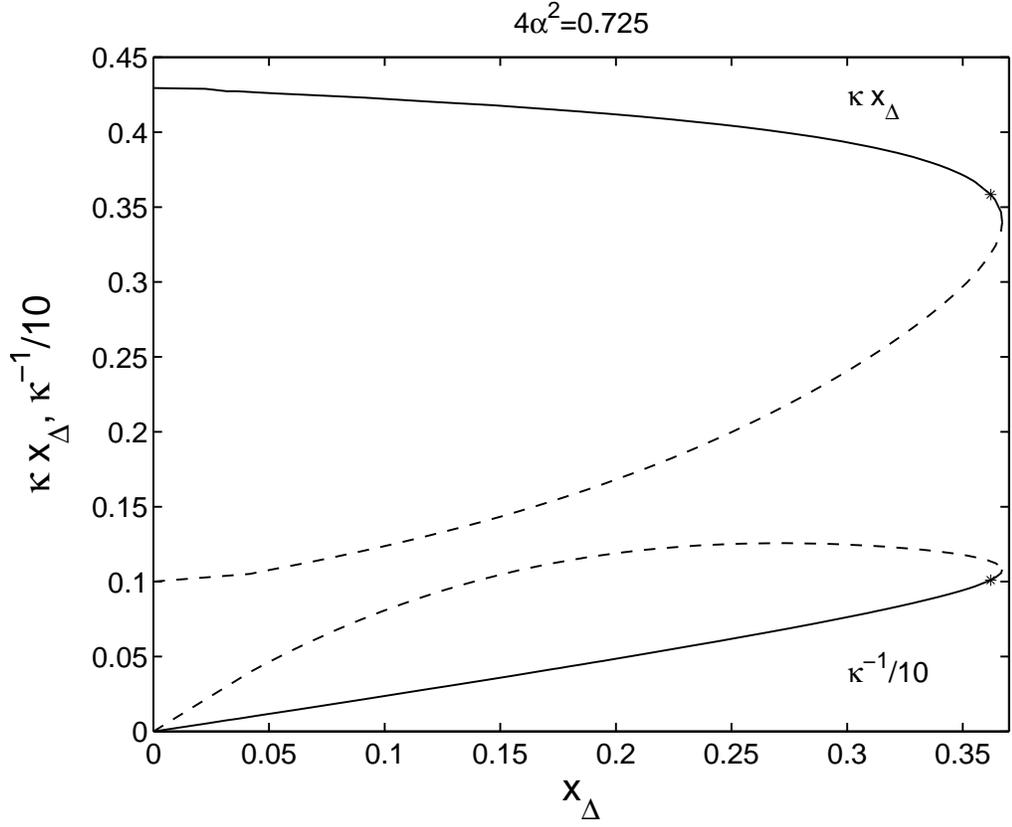}}
\caption{
Same as Fig.~1 for the inverse (rescaled) surface gravity $\kappa^{-1}$
and the integrand of the horizon mass $\kappa x_\Delta$.
(The value of $\kappa x_\Delta$ for $x_\Delta \rightarrow 0$
is extrapolated.)
}
\end{figure}

\newpage
\begin{figure}[h!]\centering
\epsfysize=11cm\mbox{\epsffile{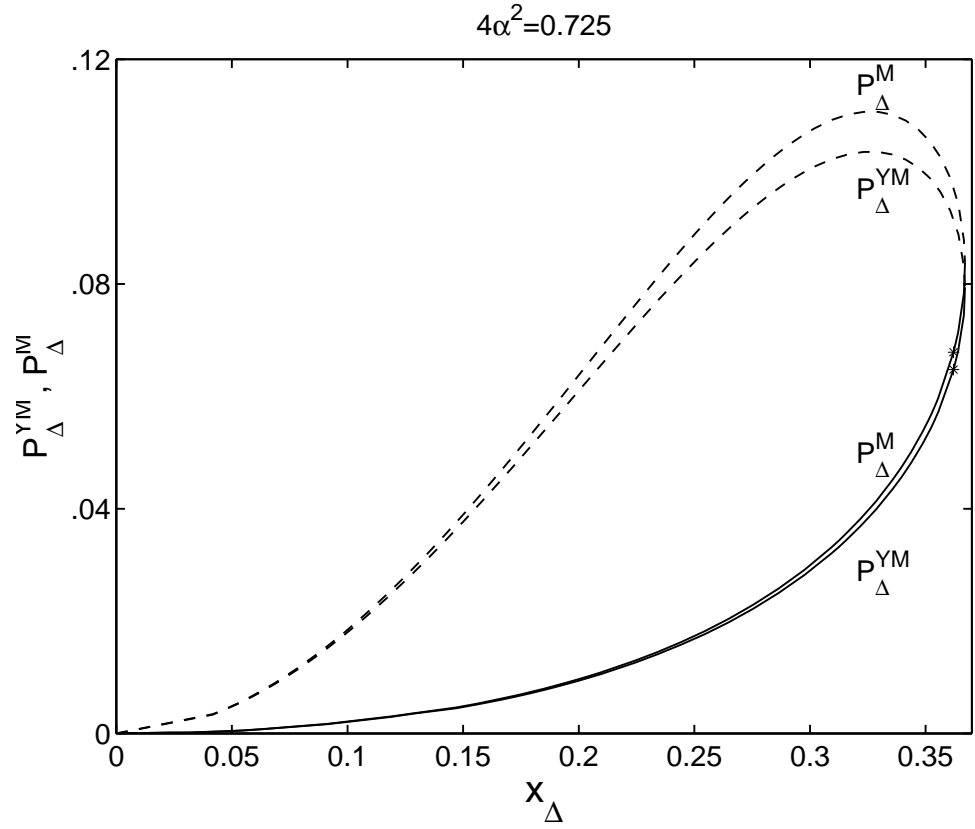}}
\caption{
Same as Fig.~1 for the
magnetic non-abelian charge of the horizon $P_\Delta^{\rm YM}$
and the magnetic abelian charge of the horizon $P_\Delta^{\rm M}$.
}
\end{figure}

\end{document}